\documentclass[fleqn,usenatbib]{mnras}

\usepackage{newtxtext,newtxmath,CJK,tablefootnote}
\usepackage[table,xcdraw]{xcolor}

\usepackage[T1]{fontenc}

\DeclareRobustCommand{\VAN}[3]{#2}
\let\VANthebibliography\thebibliography
\def\thebibliography{\DeclareRobustCommand{\VAN}[3]{##3}\VANthebibliography}

\usepackage{graphicx}	
\usepackage{amsmath}	

\title[The 2022 Encounter of the Material from 73P]{The 2022 Encounter of the Outburst Material from Comet 73P/Schwassmann--Wachmann~3}

\author[Q.-Z. Ye \& J. Vaubaillon]{
\href{https://orcid.org/0000-0002-4838-7676}{Quanzhi Ye (叶泉志)$^{1,2}$\thanks{E-mail: qye@umd.edu}},
J\'{e}r\'{e}mie Vaubaillon$^3$
\\
$^1$Department of Astronomy, University of Maryland, College Park, MD 20742, USA\\
$^2$Center for Space Physics, Boston University, 725 Commonwealth Ave, Boston, MA 02215, USA\\
$^3$Institut de M\'ecanique C\'eleste et de Calcul des \'Eph\'em\'erides, IMCCE, Observatoire de Paris, PSL Research University,\\ CNRS,Sorbonne Universit\'es, UPMC Univ Paris 06, Univ. Lille, 77 Av. Denfert-Rochereau, F-75014 Paris, France. \\
}

\date{Accepted XXX. Received YYY; in original form ZZZ}

\pubyear{2022}

\begin{document}
\label{firstpage}
\begin{CJK*}{UTF8}{gbsn}
\pagerange{\pageref{firstpage}--\pageref{lastpage}}
\maketitle
\end{CJK*}

\begin{abstract}
The encounter of the meteoric material from 73P/Schmassmann--Wachmann~3 produced during the comet's 1995 outburst in May 2022 provides a rare and valuable opportunity to understand a fragmenting comet. Here we explore various ejection configurations and their impact on the meteor outburst detected in the early hours of UT 2022 May 31. We show that the dust must have been ejected $\sim4$ to $5\times$ faster than calculated by water-ice sublimation model to best match the observed meteor activity. As only a small subset of particles with a narrow range of cross-section is expected to have reached the Earth, the large spread of meteor brightness likely indicates the presence of large but porous meteoroids in the trail. Other effects such as an enhanced lunar sodium tail and a visible glow from the meteoroid trail may have also occurred during the encounter.
\end{abstract}




\section{Introduction}

Most meteor showers originate from Earth's crossing of cometary dust trails, providing means to sample cometary material without leaving Earth's orbit. Observation of meteor showers provide a wealth of information about the recent (10--100s orbits) evolutionary history of their parents. In particular, concurrent telescopic- and meteoric- observation of cometary dust from their release at the parent to the manifestation as meteor shower at the Earth can provide a detailed picture of the evolution of the comet. One of the best examples is the disintegration of comet 3D/Biela and its manifestation as the Andromedid meteor shower in the 19th century \citep[][\S~15]{Jenniskens2006}. However, such occasion is extremely rare since most comets are only known in the era of sky surveys, and that planetary dynamics only bring a handful of dust trails to the Earth.

Jupiter-family comet (JFC) 73P/Schwassmann--Wachmann~3 is the parent comet of the $\tau$-Herculids, a meteor shower of which the activity is nearly absent in most years besides a strong outburst in 1930 \citep{Nakamura1930}. 73P experienced a major outburst in September 1995 and has undergone a series of disintegration events ever since, producing more than a hundred meter-class or larger fragments in the process \citep[e.g.][]{Crovisier1996, Ishiguro2009, Reach2009}. Independent investigations by \citet{Luthen2001}, \citet{Horii2008} and \citet[][\S~4]{Rao2021} showed that the 1995 ejecta would pass the Earth by only 0.0004~au on 2022 May 31 which may bring elevated meteor activity or even a storm. On the other hand, \citet{Wiegert2005} and alternative calculation by \citet[][\S~3]{Rao2021} show no close encounter with the 1995 ejecta in 2022. As pointed out by \citet{Rao2021}, the difference lies in the assumption of ejection condition: the standard \citet{Whipple1951} model predicts no encounter, while more powerful ejection with a speed that is a few times higher predicts a close encounter. Additional explorations by one of us (JV) shows that it only requires an ejection speed of $2.5\times$ the Whipple model to bring millimeter-class meteoroids to the Earth\footnote{\url{https://www.imcce.fr/recherche/campagnes-observations/meteors/2022the}, accessed 2022 May 2.}.

Increased activity of the $\tau$-Herculid meteor shower has been observed on 2022 May 30--31 but only reached $\sim1/50$ of storm level \citep{Jenniskens2022, Ogawa2022, Vida2022}. Since the encounter is apparently highly sensitive to ejection condition, we are interested in understanding how different configuration of dust trail can affect the visibility of the meteor outburst, as this is important for the interpretation of the meteor observation regardless of the outcome. Here, we present a suite of models constructed using available observational evidence before the meteor outburst and examine their implication on the visibility and intensity of the outburst.

\section{Modeling}

The aforementioned researchers made a number of different, but largely compatible assumptions in their works. \citet{Luthen2001}, \citet{Horii2008} and \citet{Rao2021} all assumed ejection at perihelion passage and tested a range of ejection speeds; \citet{Wiegert2005} simulated a suite of dust particles ejected within water-ice sublimation distance ($\sim3$~au) and adopted a thermophysical model to calculate ejection speeds, assuming sublimation of pure water-ice. All these approaches (or their variants) have been successfully applied on the predictions of previous meteor outbursts \citep[see, e.g.][and the reference therein]{Vaubaillon2019}.

The case of 73P, however, is complicated due to its extensive fragmentationary history. Spitzer observation of 73P's debris trail shows that the ejection speed is likely $2\times$ higher than predicted by classical pure-ice sublimation model \citep{Vaubaillon2010}. Fragment C, a major component of the 73P fragment stream, was found to be highly active at the edge of the water-ice sublimation distance \citep{Toth2005}. Activity or even significant outburst of JFC nuclei beyond the water-ice sublimation distance is unusual, but is far from uncommon \citep{Kelley2013,Ye2019}.

We constructed two models with slightly different assumptions, as describe in the following.

The first model assumes a nucleus that is active over its entire surface along the entire orbit. The purpose is to probe any meteor activity caused by distant activity of the comet. Simulated particles are released from 1995 September 12 \citep[the approximate onset of the fragmentation][]{Crovisier1996} to 1999 October 21 (an arbitrary date that is beyond the aphelion date of the comet, which occurred on 1998 May 27) to cover the aphelion passage of the comet. We test the classical pure-ice model \citep[used here is the][model]{Whipple1951} as well as the same model with speeds multiplied by $2\times$, $3\times$, $4\times$, and $5\times$, corresponding to approximate ejection speeds of $800$, $1200$, $1600$, and $2000$~m/s of $1~\micron$ grains at 1~au. (We note that the Spitzer observation reported by \citet{Vaubaillon2010} is consistent with the $2\times$ model, hence these are effectively $0.5\times$, $1\times$, $1.5\times$, $2\times$ the Spitzer observation.) Particles being simulated have $\beta<0.03$\footnote{$\beta$ is the ratio between radiation pressure and solar gravity and can be calculated using $\beta=5.74\times10^{-4}/(\rho_\mathrm{d} a)$ where $\rho_\mathrm{d}$, $a$ are bulk density (in kg/m$^3$) and radius (in m) of a spherical particle \citep[cf.][]{Burns1979}. $\beta=0.03$ roughly translates to particles with a radius of 10~\micron. In this paper, we try to stick to $\beta$ instead of $a$ when discussing simulation results to eliminate the need to refer to bulk density and particle shape as both quantities are poorly constrained.} which includes meteoroids from the optical down to the high-power large-aperture (HPLA) radar regime, as well as a differential size distribution of $N(a) \propto a^{-3.7}$ following the measurement reported by \citet{Vaubaillon2010}. The amount of dust generated is inferred from the fact that the comet peaked at a visual magnitude $V\sim6$ during its outburst in 1995. Since the calculation by \citet{Wiegert2005} already showed that the treatment of different fragment does not significantly influence the outcome, we exclusively use the orbit solution JPL K012/14\footnote{Available from the JPL Small Body Database, \url{https://ssd.jpl.nasa.gov/tools/sbdb_lookup.html}.} of the main body of 73P as the parent particle in our simulation. This solution was computed using the observation from 1994 December 28 to 2000 November 4, which grossly overlap with our assumed window of activity. Particles are then integrated using a Bulirsh--Stoer integrator, with the consideration of the gravitational perturbations from the eight major planets (the Earth-Moon system is represented by a single particle at the barycenter of the two bodies), radiation pressure, as well as the Poynting-Robertson drag. We then follow the procedure described in \citet[][\S~2]{Ye2016} to calculate the visibility and flux of meteor outbursts at the Earth.

The second model uses a more ``conventional'' approach described by \citet{Vaubaillon2005b} and \citet{Vaubaillon2005a}, which based on a hydrodynamical ejection model \citep{Crifo1997} and assumes ejection at $<3$~au over the sunlit hemisphere. This model only considers meteoroids with $\beta<0.006$ (roughly translates to particles with a radius of 100~\micron) which are responsible for optical and typical radar meteoroids. The only ``tweak'' to the ejection model is that we tested the $2\times$ and $2.5\times$ ejection speed as suggested from cursory attempts mentioned in \S~1.

Both models assume a nucleus diameter and bulk density of 1~km and $1000~\mathrm{kg~m^{-3}}$.

\section{Results}

Despite difference in assumptions, the two models produce similar results. These results are comparable to predictions made by other researchers, summarized in Table~\ref{tbl:mdl}.

\begin{table*}
    \centering
    \caption{Summary of predictions made by various modelers regarding the 2022 encounter of the 1995 dust trail together with the observational results. The predicted and observed geocentric speeds are between 11--12~km/s. $\beta<0.03$, $\beta<0.006$ and $\beta<0.0003$ roughly translate to dust grains larger than 10, 100~\micron and 1~mm. ZHR is only given for the cases of $\beta<0.0003$ since the concept of ZHR is only applicable to optical, millimeter-class meteoroids.}
    \label{tbl:mdl}
    \begin{tabular}{ccccc}
    \hline
    \hline
    Model/observation & Scenario & Peak time (UT) & Radiant ($\alpha$, $\delta$) & Note \\
    \hline
    This work -- model 1 & Nominal speed, $\beta<0.03$ & 2022 May 31 03:26 & $207.2^\circ$, $+27.7^\circ$ & Encounter with $\beta<0.003$ dust only \\
    .. & $2\times$ speed, $\beta<0.03$ & 2022 May 31 03:53 & $207.1^\circ$, $+27.8^\circ$ & Encounter with $\beta<0.003$ dust only \\
    .. & $3\times$ speed, $\beta<0.0003$ & 2022 May 31 04:30 & $209.5^\circ$, $+28.1^\circ$ & Peak ZHR=47, FWHM=1.1~hr \\
    .. & $4\times$ speed, $\beta<0.0003$ & 2022 May 31 04:44 & $209.5^\circ$, $+28.1^\circ$ & Peak ZHR=76, FWHM=2.0~hr \\
    .. & $5\times$ speed, $\beta<0.0003$ & 2022 May 31 04:34 & $209.5^\circ$, $+28.1^\circ$ & Peak ZHR=70, FWHM=4.2~hr \\
    This work -- model 2 & $2\times$ speed, $\beta<0.006$ & - & - & No encounter \\
    .. & $2.5\times$ speed, $\beta<0.006$ & 2022 May 31 05:01 & $209.4^\circ$, $+28.3^\circ$ & \\
    \citet{Rao2021} & - & 2022 May 31 05:59 & $210.17^\circ$, $+25.03^\circ$ & \\
    \citet{Horii2008} & - & 2022 May 31 04:59 & $209.48^\circ$, $+28.13^\circ$ & \\
    \citet{Jenniskens2006} & - & 2022 May 31 05:17 & - & \\
    \citet{Wiegert2005} & - & - & - & No encounter \\
    \citet{Luthen2001} & - & 2022 May 31 04:55 & $205.40^\circ$, $+29.20^\circ$ & \\
    M. Maslov$^a$ & - & 2022 May 31 05:15 & $209.5^\circ$, $+28.0^\circ$ & ZHR=600+ \\ 
    M. Sato$^b$ & - & 2022 May 31 05:04 & - & - \\ 
    \hline
    CAMS$^c$ & & 2022 May 31 04:42 & $209.17^\circ$, $+28.21^\circ$ & FWHM=3.5~hr \\
    GMN$^d$ & & 2022 May 31 04:15 & $208.6^\circ$, $+27.7^\circ$ & Peak ZHR=22, FWHM$\sim4$~hr \\
    IPRMO$^e$ & & 2022 May 31 03:30--04:30 & - & Peak ZHR=34, FWHM=9~hr \\
    IMO$^f$ & & 2022 May 31 04:58 & - & Peak ZHR=50, FWHM=4~hr \\
    \hline
    \hline
    \multicolumn{5}{l}{\footnotesize$^a$ \url{http://feraj.ru/Radiants/Predictions/1901-2100eng/73p-ids1901-2100predeng.html}, accessed 2022 May 15.}\\
    \multicolumn{5}{l}{\footnotesize$^b$ International Meteor Organization 2022 Meteor Shower Calendar, \url{https://www.imo.net/files/meteor-shower/cal2022.pdf}, accessed 2022 May 15.}\\
    \multicolumn{5}{l}{\footnotesize$^c$ From \citet{Jenniskens2022}.}\\
    \multicolumn{5}{l}{\footnotesize$^d$ From \citet{Vida2022}.}\\
    \multicolumn{5}{l}{\footnotesize$^e$ From \citet{Ogawa2022}.}\\
    \multicolumn{5}{l}{\footnotesize$^f$ \url{https://www.imo.net/members/imo_live_shower?shower=TAH&year=2022}, accessed 2022 June 19.}\\
    \end{tabular}
\end{table*}

\begin{figure*}
\begin{center}
\includegraphics{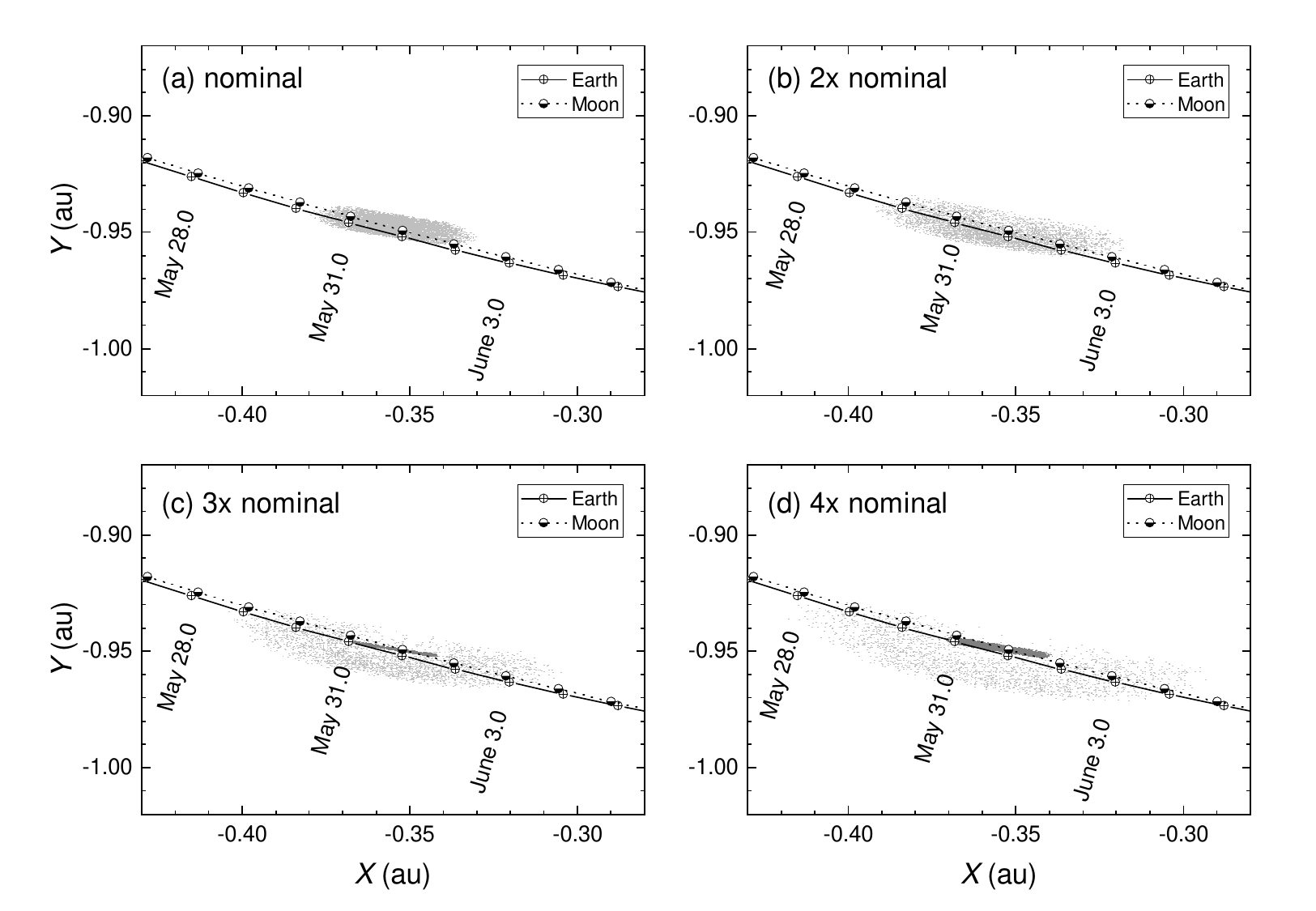}
\caption{Footprint of the 1995 ejecta onto Earth's orbit on 2022 May 31 under different scenarios in model 1. Meteoroids plotted are the ones that come within 1 lunar distance from the ecliptic plane. The points in light grey represent meteoroids between $\beta=0.0003$ and $0.03$ (roughly between 10~\micron and 1~mm), while the points in dark grey (only present in panels c and d) represent meteoroids with $\beta<0.0003$ (roughly $>1$~mm). All models adopt the same dust production rate of the comet and simulate the same number of particles. The lower spatial number density seen in the $4\times$ nominal model reflects the fact that the material is more spread out at higher ejection speed.\label{fig:footprint}}
\end{center}
\end{figure*}

Figure~\ref{fig:footprint} shows the footprint of the simulated particles under different scenarios in model 1 on UT 2022 May 31.0. All these scenarios confirm a prolonged encounter of the debris trail. The smaller, sub-millimeter-class meteoroids tend to arrive about an hour earlier than the larger, millimeter-class ones.

We also confirm that while sub-millimeter-class meteoroids will reach the Earth in all scenarios, both models show that the millimeter-class meteoroids (responsible for meteors in optical regime) only reach the Earth when the ejection speed is at least $2.5\times$ to $2.75\times$ the nominal scenario. Model 1 shows that the flux of meteoroids of 10~\micron~or larger is on the order of $10^2~\mathrm{km^{-2}~hr^{-1}}$ across all scenarios, dominated by meteoroids smaller than 100~\micron. The flux of millimeter-class meteoroids, if they do reach the Earth, is around the order of $10^{-2}~\mathrm{km^{-2}~hr^{-1}}$, equivalent to a ZHR of $\sim50$. (We note that the flux does not increase proportionally with an increasing ejection speed, as material reaching the Earth is more spread out, leading to lower volume density.) We also find that the meteoroids reaching the Earth were exclusively ejected between the outburst onset (early September 1995) to February 1996 (when the comet was at $\sim2.0$~au from the Sun in its outbound leg). This implies that meteoroids ejected from distant activity of 73P, including those from the presumed disruption of transient fragments (e.g. fragments A, D, and possibly E which split off from B some time between 1995 and 2001) will not reach the Earth in 2022.

\section{Discussion}

Our simulation shows that millimeter-class meteoroids responsible for optical meteors can only reach the Earth when the ejection speed is at least $40\%$ higher than suggested by the Spitzer observation, or $\sim3\times$ higher than the speed calculated by \citet{Whipple1951}'s model. This must have been the case since significant activity has been detected. The observed peak of activity, radiant, ZHR and full-width-half-maximum (FWHM) of the activity profile are in general agreement with the models in Table~\ref{tbl:mdl}. In particular, the radio observations by the International Project for Radio Meteor Observations (IPRMO), which is sensitive to smaller meteoroids, reported a peak that is slightly earlier and more prolonged than measured in optical, as expected by the models. The $4\times$- and $5\times$-Whipple-speed scenarios appear to provide the best match to the observed ZHR and FWHM \citep{Jenniskens2022,Vida2022}.

An noteworthy feature of the outburst is the detection of bright meteors by many observers. Models predict that the Earth would only intercept meteoroids within a narrow range of $0.0002<\beta<0.0006$, since more massive (smaller $\beta$) meteoroids would be too slow to reach the Earth. Assuming a compact spherical grain, a 0-mag $\tau$-Herculid meteor would need a meteoroid of a few centimeter in radius to produce \citep[cf.][Fig. 1]{Ye2016}. However, a typical cm-class meteoroid would require an ejection speed more than an order of magnitude higher than the speed calculated by Whipple's model (and thus, a few times higher than the $5\times$ model described above) to reach the Earth, which is improbable. Hence, these bright meteors likely indicate the presence of centimeter-class, porous dust aggregates that dynamically behave like compact millimeter-class meteoroids. This can also explain the observed fragility of the meteoroids from the 1995 trail but not other trails \citep{Vida2022, Ye2022}.

We also note that the Moon is slightly closer to the center of the trail. As a result, meteor activity (or rather, meteoroid bombardment) lasts longer on the Moon compared to the Earth. However, the dominance of $<100~\micron$ meteoroids as well as the low arrival speed seem to suggest that few of these events can be detected as lunar flashes from the ground. The dominance of $<100~\micron$ meteoroids, on the other hand, may provide insight into the driver of the lunar sodium tail which appears to be correlated with the rate of sporadic meteors that are of similar sizes \citep{Baumgardner2021}.

\begin{figure*}
\begin{center}
\includegraphics[width=0.8\textwidth]{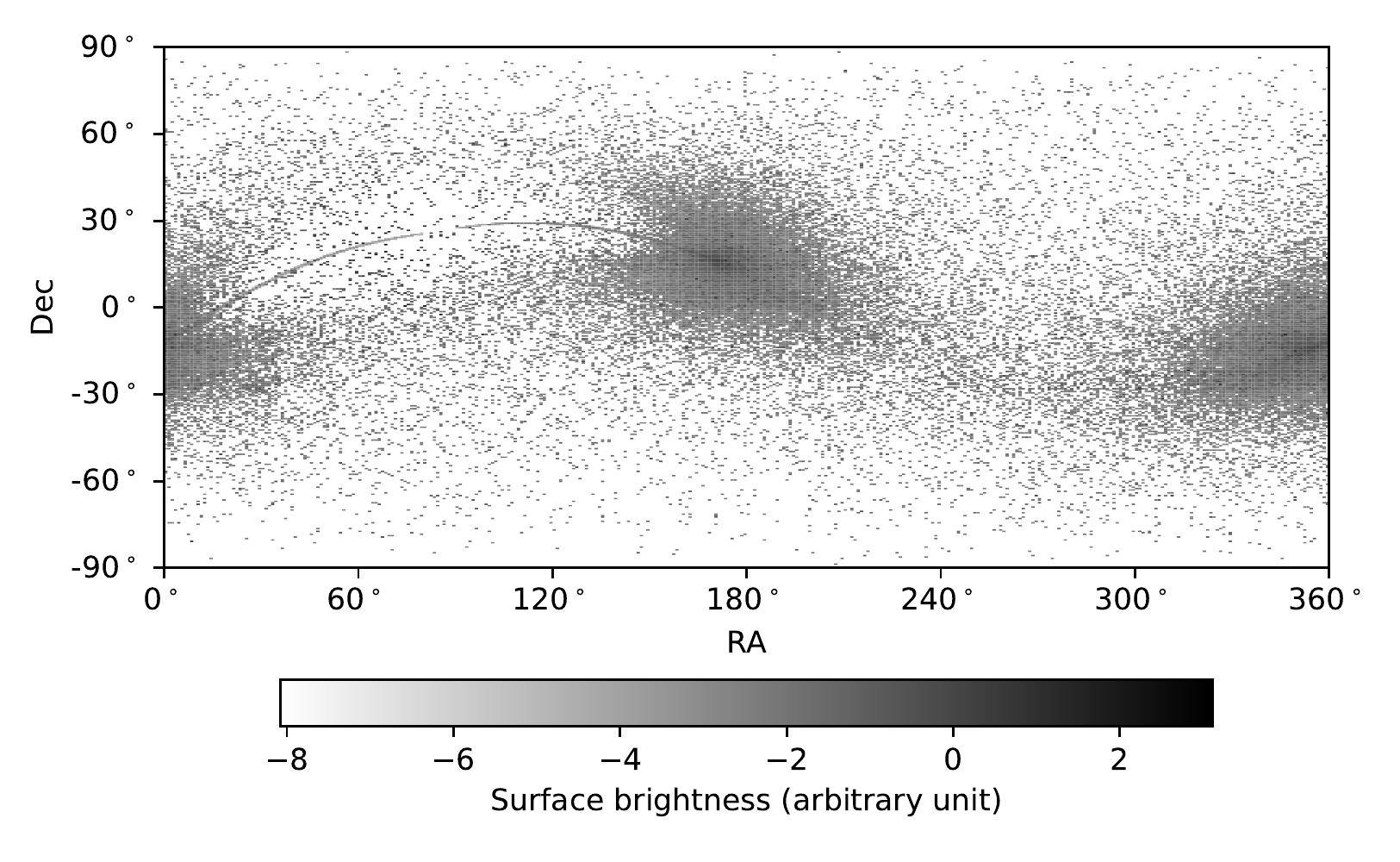}
\caption{On-sky distribution and surface brightness (in arbitrary unit) of the 1995 trail as viewed from geocenter at UT 04~h, 2022 May 31. \label{fig:cloud}}
\end{center}
\end{figure*}

Besides manifesting into detectable meteor activities, the meteoroid trail may be close enough to the Earth so that it can be detected as a faint glow in the sky. \citet{Nakamura2000} reported a similar glow detected during the 1998 Leonid meteor storm and derived a number density of $10^{-10}~\mathrm{m^{-3}}$ assuming 10-\micron-class meteoroids. Figure~\ref{fig:cloud} shows the on-sky projection of the 1995 trail at 04~h UT, 2022 May 31, as viewed from the Earth. The trail projects into two ``lobes'' as the Earth passes through it, with one lobe centers at about $\alpha=170^\circ$, $\delta=+20^\circ$ in the constellation of Leo, while the other lobe centers at $\alpha=355^\circ$, $\delta=-15^\circ$ in the constellation of Equuleus. Based on the simulation results, we estimate that the number density for the trail is in the order of $10^{-12}$ to $10^{-11}~\mathrm{m^{-3}}$, lower than the 1998 Leonid trail but perhaps possible to detect.

\section{Conclusion}

We confirm that the Earth did pass very close to the debris trail produced by comet 73P/Schwassmann--Wachmann~3 during its large outburst and subsequent fragmentation in 1995 around 2022 May 31. The ejection speed of the 1995 event must have been $40\%$ higher than previously constrained by Spitzer observation, or $\sim3\times$ higher than value calculated by Whipple's model, in order to explain the arrival of the swarm of millimeter-class meteoroids which have been detected near the predicted time of meteor activity. The observed activity profile of the meteor outburst is best matched by the models that assume $4\times$ to $5\times$ of the speed calculated by Whipple's model.

The center of the meteoroid trail will pass a few tenths of an au in the sunward direction of the Earth during New Moon. Given the possibly high number density of the debris trail, effects such as an enhanced lunar sodium tail may also occur. The faint glow of the trail may also be visible during the close approach.

We also note the crossing of the older 1892 and 1897 trails might have produce two separate meteor outbursts around UT~16~h of May 30 and 10~h of May 31 \citep{Wiegert2005}, as reported by \citet{Ogawa2022} and \citet{Ye2022}. Given that 73P was only discovered in 1930 and has only been widely observed since the 1990s, investigations of these two meteor outbursts, along with the one produced by the 1995 trail as studied in this paper, will provide useful information and constraints about the history of the comet.

\section*{Acknowledgements}

We thank Galina Ryabova for her careful review and comments, as well as Mike Kelley and Matthew Knight for discussions.


\section*{Data Availability}

All data are incorporated into the article.



\bibliographystyle{mnras}
\bibliography{sample631} 





\bsp	
\label{lastpage}
\end{document}